\documentstyle[eqsecnum,aps]{revtex}
\begin{document}

\draft
\title{Relativistic Quantum Field Theory with a Physical State Vector}    
\author{Bernd A. Berg}
\address{
 Department of Physics,
 Florida State University\\
 Tallahassee, FL 32301, USA}
\maketitle

\begin{abstract}
Evolution of a physical quantum state vector is described as governed by 
two distinct physical laws: Continuous, unitary time evolution and a 
relativistically covariant reduction process. In previous literature, 
it was concluded that a relativistically satisfactory version of the
collapse postulate is in contradiction with physical measurements of a 
non-local state history. Here it is shown that such measurements 
are excluded when reduction is formulated as a physical process and
the measurement devices are included as part of the state vector.

\noindent
SERVER quant-ph/9807046;
\end{abstract}
\vskip.2in\indent

\section{Introduction}

Relativistic Quantum Field Theory (QFT) relies on a number of 
interpretational rules which allow to make measurable predictions. 
In the non-relativistic limit of Quantum Theory (QT) a state vector can 
be defined which evolves according to two laws: (a)~Unitary, 
continuous time evolution in agreement with the non-relativistic equations 
of motion and (b) the state vector reduction or collapse postulate,
which dictates how the state vector changes as a result 
of measurement. Together, they allow to define a unique state vector 
history. In the following the notation Physical State Vector (PSV) 
is used for a state vector which, in its coordinate space 
representation, is well defined at all spacetime points.

For the relativistic theory the definition of a PSV faces 
difficulties~\cite{Bl67} due to the fact that spacelike measurements 
which are instantaneous in one inertial frame are no longer 
instantaneous in another. Collapse has to take place in a covariant
fashion. This requirement restricts us to using light cone 
sections. It turns out that the Forward Light Cone (FLC) is insufficient, 
as follows most stringent from experiments (see~\cite{exp} and references 
therein) performed in connection with measuring violations of 
Bell's~\cite{Be64} inequalities. Indeed, a prescription for reduction on 
the Backward Light Cone (BLC) was already proposed in an early 
paper~\cite{HeKr70} on the subject. But, subsequently Aharonov and 
Albert~\cite{AhAl80} (AA) concluded that a relativistically satisfactory 
version of the collapse postulate cannot be found at all. Their analysis 
relies on constructing {\it local devices} which allow to monitor a 
{\it non-local history}. As stressed by them, the empirical meaning of 
their procedure is crucial. Otherwise it would be without physical content 
to assert that the system is in an eigenstate of some non-local operator 
and, being in such a state, excluding collapse on the BLC. In summary, 
AA state that the relativistic theory still has the 
capacity to predict probabilities, but lacks the capacity to define 
a consistent PSV.
\medskip

It is one purpose of the present paper to show that measurements of the 
Aharonov-Albert type are ruled out, once the measurement device is
considered to be part of the PSV and reduction is understood as
a physical process, distinct from unitary and continuous time evolution.
In section II.A the non-relativistic PSV evolution is reviewed,
followed by introducing its relativistic generalization. Section II.B
introduces an unconventional formulation of the reduction process. Namely,
each reduction is attributed to the action of a detector. For the
non-relativistic limit this is shown to agree with the conventional 
formulation. The full relativistic spacetime picture is developed in 
section~II.C and our rules for reduction are distinct from those of 
ref.\cite{HeKr70}. It is shown in section~II.D that AA
measurements become impossible within the developed framework.
\medskip

To give some illustrations, section~III deals first with spacelike
measurement of charge for a particle which is in a superposition of
two waves localized at spacelike positions. The introduction of
AA-like devices (for a precise definition see there) does not lead to 
inconsistencies. Section~II.B considers spacelike measurements 
on the entangled singlet state of two distinguishable spin~$1\over 2$ 
particles. This example, also discussed by Aharonov and Albert, is for 
various features of interest the technically simplest illustration. 
In particular, it turns out that the remains of the AA idea are
still sufficient to show inconsistency for the proposal by
Hellwig and Kraus~\cite{HeKr70}. In section III.C measurement is
briefly discussed for the Greenberger-Horne-Zeilinger (GHZ)
state~\cite{GHZ89,Me91}, mainly to give an example where three
detectors at spacelike positions are involved.
Conclusion are drawn in the final section~IV. 
\medskip

\section{Physical State Vector and Reduction}

We deal with states in the Schr\"odinger picture. The question may
arise, whether statistical mixtures ought better to be considered. 
Let us first have a look at classical physics. There,
the existence of a single state is guaranteed by definition and
statistics is an essential tool in view of the practical impossibility 
to control a large number initial conditions. Here I take a similar
stance on QT, I assume that a PSV of the world $|\Psi_W\rangle$ exists, 
although a complete set of measurements to determine the initial state 
cannot be performed. The essential feature of QT is that even complete 
knowledge of the state does only allow stochastic predictions. Introducing 
a density matrix approach at this point, where we are dealing with the 
fundamental nature of QT, may more obscure than enlighten this central 
issue. Whether a state or a density matrix reflects the underlying 
fundamental physics correctly, either case is an assumption, and Newton's 
First Rule for Reasoning in Philosophy decides the matter in favor of the 
state.
\medskip

\subsection{General state vector evolution}

In the following a microscopic object (photon, electron, proton, etc.), 
as studied in many QT applications, is called Quantum Object (QO). Let 
us define
\begin{equation} \label{tlimits}
 t_+ = \lim_{\epsilon\to 0, \epsilon>0} (t+\epsilon) ~~{\rm and}~~
 t_- = \lim_{\epsilon\to 0, \epsilon>0} (t-\epsilon)\
\end{equation}
in the sense that whenever these symbols are used the indicated limits
have to be taken outside of all other operations.
In the non-relativistic QT state vectors of QOs $|\psi (t)\rangle$
are defined on the instantaneous hyperplane of $4d$ spacetime and
evolve according to two laws:

\begin{description}

\item{(a)} Unitary, causal time evolution $(t_{1-}>t_{0+})$
\begin{equation} \label{law_a}
 |\psi (t_{1-})\rangle = U(t_{1-}-t_{0+})\, |\psi (t_{0+})\rangle 
 ~~{\rm with}~~ U(t) = \exp (-i\,H\,t ) 
\end{equation}
when no measurements are carried out.

\item{(b)} An instantaneous, stochastic transformation (state vector 
reduction or collapse)
\begin{equation} \label{law_b}
 |\psi (t_{1-})\rangle \to |\psi (t_{1+}) \rangle
\end{equation}
when a measurement is carried out at time $t_1$.

\end{description}

The initial state $|\psi(t_{0+})\rangle$ can, in principle, uniquely be
specified by measuring a complete, commuting set of observables. Let 
${\cal A}$ represent operators corresponding to such a set of observables 
and assume that the possible results of their measurements are given
by $a_n$, $(n=0,1,2, ...)$, where each $a_n$ represents an
entire set of eigenvalues and, hence, specifies a unique state 
$|\psi_n\rangle$ for which
\begin{equation} \label{complete}
{\cal A}\, |\psi_n\rangle = a_n\, |\psi_n\rangle
\end{equation}
holds. Assume, the operators of ${\cal A}$ are measured at time $t_A>t_0$, 
then QT predicts that $a_n$ is found with probability
\begin{equation} \label{Prob} 
P_n = | \langle \psi_n | \psi (t_{A-}) \rangle |^2\ .
\end{equation} 
Given that the particular set of eigenvalues $a_n$ is observed at time
$t_A$, the state $|\psi (t_{A+})\rangle$ of (\ref{law_b}) is 
$|\psi (t_{A+})\rangle = |\psi_n\rangle$. In this way a series of complete
measurements ${\cal A},\, {\cal B},\, {\cal C},\, ...$ at times 
$t_A<t_B<t_C<\, ...$ (where
each capital letter represents a complete set of operators) defines a
{\it physical state} history $|\psi(t)\rangle$. This state history can
be monitored by performing non-demolition experiments, which measure
the last complete set of operators again, before the next reduction.
For instance, for time $t$ in the range $t_A < t < t_B$ we would measure
the observables $U(t-t_A)\,A\,U^{-1}(t-t_A)$. This reproduces the result
$a_n$ with certainty and without disturbing the state
$|\psi (t)\rangle = |\psi_n (t-t_A)\rangle = U(t-t_A)\,|\psi_n\rangle$.
\medskip

The idea of a PSV has considerable intuitive appeal and should not be given 
up easily. Here I embark on a starting point which is kind of opposite to 
the one of ref.\cite{AhAl80}. Assuming the existence of 
$|\Psi_W\rangle$, a distinct understanding of measurements is developed. As
explained in the next subsections, each reduction is attributed to a detector
and new restrictions emerge, because every detector has to be part of the PSV. 
While the non-relativistic limit is found to agree with the scenario sketched 
above, AA measurements become ruled out, and a consistent PSV may exist.
\medskip

First, we need some notation. In the following we call two spacetime 
points $x_0=(ct_0,\vec{x}_0)$ and $x_1=(ct_1,\vec{x}_1)$ spacelike when
\begin{equation} \label{spacelike}
(x_0-x_1)^2 = c^2\,(t_0-t_1)^2 - (\vec{x}_0-\vec{x_1})^2 \le 0 
\end{equation}
holds. In case of the equal sign, we assume that $x_0$ and $x_1$ are defined
as the limit of points 
$$ x_0 = \lim_{\delta\to 0,\delta>0} x_{0,\delta} ~~{\rm and}~~
   x_1 = \lim_{\delta\to 0,\delta>0} x_{1,\delta} ~~{\rm such\ that}~~
  (x_{0,\delta}-x_{1,\delta})^2 < 0~~{\rm holds\ for\ all}~~\delta>0\,.$$
A Lorentz Covariant Spacelike Hypersurface (LCSH)
is now defined to be a hypersurface $S$ which fulfills two conditions: 
(1)~Equation~(\ref{spacelike}) holds for any two points $x_0\in S$ and 
$x_1\in S$. (2)~To insure Lorentz covariance of the collapse process, 
$S$ is build from light cone
sections. Let $S_0$ and $S_1$ be LCSHs. The hypersurface $S_1$ is said to be 
in the {\it future} of $S_0$, when for any two points $x_0\in S$ and 
$x_1\in S_1$ with $\vec{x}_0=\vec{x}_1$ we have $t_0\le t_1$, and there 
exist some such points with $t_0 < t_1$. 
Similarly single spacetime points or sets of such points are defined to 
be in the future (or past) of a LCSH. Finally, for each LCSH $S$ 
hypersurfaces $S_+$
and $S_-$ are defined by taking the time component of all their spacetime 
points in accordance with the limits defined by equation~(\ref{tlimits}).
Assume a state vector is initially defined on some LCSH $S_{0+}$. Its 
relativistic evolution follows then rules similar to the laws (a) and (b):

\begin{description}

\item{(a$'$)} Unitary and causal evolution to some LCSH $S_{1-}$ in the
future of $S_{0+}$
\begin{equation} \label{law_a'} 
|\Psi (S_{1-})\rangle\ =\ U(S_{1-},S_{0+})\ |\Psi (S_{0+})\rangle
\end{equation}
when no reduction happens on any hypersurface in-between $S_{1-}$ and $S_{0+}$.

\item{(b$'$)} A stochastic transformation
\begin{equation} \label{law_b'} 
|\Psi (S_{1-})\rangle\ \to\ |\Psi (S_{1+})\rangle
\end{equation}
describes reduction on $S_1$. From $S_{1+}$ the state vector evolves to 
a next LCSH $S_{2-}$ by unitary and causal evolution. On $S_2$ reduction 
takes place, and so on. 

\end{description}

The process is graphically depicted in figure~1. A light cone section 
becomes added to the initial LCSH $S_0$ such that the created
surface $S_1$ is in the future of $S_0$. On $S_1$ reduction takes place
from $S_{1-}$ to $S_{1+}$. The $S_{0\pm}$ and $S_{1\pm}$ surfaces are
not explicitly indicated in figure~1, their locations are obvious.
In the next subsection we consider the 
reduction~(b$'$) in detail and come to a process where the
addition of a single new light cone section (not spelled out in rule
(a$'$)) is typical.  By reasons explained then, the definition of 
reduction in (b$'$) is not entirely identical with the one of (b). Namely,
some measurements will constitute reductions and others not. This is just
a notational issue. The physical content of (a) and (b) is fully recovered
in the non-relativistic limit of (a$'$) and (b$'$). In particular, for the
speed of light $c\to\infty$ the LCSH of figure~1 becomes the instantaneous 
hyperplane.




\subsection{Measurement and reduction}

From the fact that experimental measurements are made, it follows that it 
is possible to construct bound states $|\Psi_D\rangle$, called detectors,
which have the ability to perform them. Eventually the detector may 
project some quantum state on an eigenstate of certain operators, or 
onto a state out of the space of eigenstates in case that the eigenvalue 
in question is degenerate. I proceed with a description in terms of the
PSV: A detector $|\Psi_D\rangle$ can be constructed and a QO $|\psi\rangle$
(for instance a single electron) can be prepared, such that over some 
time~\cite{time} period the factorization
\begin{equation} \label{Wfct}
|\Psi_W\rangle = |\psi\,\Psi_D\rangle\ |\Psi_R\rangle\ , 
\end{equation}
where $|\Psi_R\rangle$ describes the rest of the world, is meaningful 
in the following sense: Although, due to interactions, this factorization 
can never be exact, the evolution of 
\begin{equation} \label{Psi}
|\Psi\rangle = |\psi\,\Psi_D\rangle\ , 
\end{equation}
considered in isolation, describes nevertheless the features we are
interested in correctly and proceeds as follows. Initially there is
very little overlap between the wave functions of the QO and of the
detector, such that approximately
\begin{equation} \label{detector}
|\Psi\rangle = |\psi\rangle\ |\Psi_D\rangle\ . 
\end{equation} 
Subsequently, $|\psi\rangle$ and $|\Psi_D\rangle$ interact and this
is expressed by the notation $|\psi\,\Psi_D\rangle$, which is already
used in equations (\ref{Wfct}) and (\ref{Psi}). At some point in its 
spacetime the detector performs a transition
\begin{equation} \label{measure}
|\psi\,\Psi_D\rangle \to |\Psi^{'n}_D\rangle 
\end{equation} 
such that the (macroscopic) state $|\Psi^{'n}_D\rangle$ constitutes a 
measurement of an observable or set of observables ${\cal O}_n$, according 
to the purpose for which the detector was constructed. The measurement
(\ref{measure}) is here interpreted as a physical ability of the detector.  
Besides avoiding the word reduction at the moment and emphasizing the role 
of the detector as part of a state vector this is not distinct from 
standard QT. In the non-relativistic limit complete 
measurements~(\ref{complete}) can be performed to the extent that
it is possible to design a detector which measures the complete set
of operators~${\cal A}$ such that the state $|\Psi^{'n}_D\rangle$ 
factorizes in the form
\begin{equation} \label{factor}
|\Psi^{'n}_D\rangle = |\psi_n\rangle\ |\Psi^n_D\rangle
\end{equation} 
where $|\psi_n\rangle$ is the eigenstate of (\ref{complete}) and the
state $|\Psi^n_D\rangle$ allows us to access this information (through
the neglected interactions with $|\Psi_R\rangle$). 
\medskip

It should be noted that for real detectors the factorization (\ref{factor}) 
is the exception and not the rule. To give one example, to measure the 
$L_z$ spin component of an electron ala Stern-Gerlach one does first 
split the electron wave function by employing a magnetic field. 
Subsequently one detects the electron in one of two alternative 
branches, for instance by using a Channel Electron Multiplier (CEM) 
which actually digests the electron. We may employ just one CEM,
say in the branch corresponding to $L_z=+{1\over 2}$, and distinguish
two states of the CEM: $|\Psi_D\rangle$ when nothing happens and 
$|\Psi'_D\rangle$ when a signal current is put out. The interpretation 
of the latter case is that a $L_z=+{1\over 2}$ electron has been
detected. Further, if the detection efficiency would be 100\% and 
we could make sure by some initial measurement that one electron enters
the device, we would interpret the $|\Psi_D\rangle$ state as a measurement
of $L_z=-{1\over 2}$. In practice, one would rather count the electron
in the $L_z=-{1\over 2}$ branch with a second CEM. To stay close with what 
actually happens, we use in the following the symbol $|\Psi_D^{'n}\rangle$
to denote measurement results, with the understanding that 
factorization (\ref{factor}) is included as a special case.
\medskip

We are now ready to define reduction. For the measurement process
(\ref{measure}) we distinguish two cases:

\begin{description}

\item{(1)} $|\psi\rangle = |\psi_n\rangle$ holds for the initial QO. Then
the transformation (\ref{measure}) appears to be consistent with the 
time evolution (a$'$) and we assume that this is indeed the case.
For the just given practical example this would be when
an electron is either initially prepared in the spin $L_z=+{1\over 2}$ 
state and the resulting detector state is $|\Psi_D'\rangle$, or
when the electron is initially prepared in the $L_z=-{1\over 2}$
state and the resulting detector state is $|\Psi_D\rangle$. 

\item{(2)} $|\psi\rangle = c_n\,|\psi_n\rangle + c_{\bar{n}}\,
|\psi_{\bar{n}}\rangle$ with $|c_n|^2+|c_{\bar{n}}|^2=1$,
both $|c_n|^2>0$ and $|c_{\bar{n}}|^2>0$, and 
$\langle \psi_n | \psi_{\bar{n}} \rangle = 0$ holds for the initial
QO. Then the transformation (\ref{measure}) appears to be inconsistent
with the time evolution (a$'$). We assume that this is indeed the case 
and understand reduction (b$'$) as a physical law distinct from (a$'$). In 
the following we limit the definition of reduction strictly to the 
situation where it constitutes an interruption of the time evolution~(a$'$). 
For our practical example reduction would, for instance, take place when 
the electron is initially prepared in an $L_x$ eigenstate, say 
$L_x={1\over 2}$. The possible outcomes, each with 50\% probability, 
are then the CEM states $|\Psi_D\rangle$ versus $|\Psi'_D\rangle$, 
both somehow labelled by the superscript $n$ in equation~(\ref{measure}).

\end{description}
\medskip

Case (1) is measurement without reduction. When, in addition, also the 
factorization (\ref{factor}) holds, we have a non-demolition measurement.
Although not every measurement constitutes now a reduction, every 
detector has to be attributed the ability to perform reductions. This
follows from the fact that a detector is a device constructed for
performing measurements of some observable(s) on certain QOs. We can
then confront the detector with a QO which is prepared to have some 
component orthogonal to the measured eigenstate(s). According to the
standard rules of QT, the detector has then to make a decision between
the eigenstate(s) and the orthogonal complement, {\it i.e.} performs
a reduction in the sense just defined. 
\medskip

To move, in a finite time, from one detector state to another requires
a difference in energy distribution $\triangle E$. Assuming that 
the only driving property which constitute the ability to perform
reductions is being a boundstate confronted with a (not necessarily 
macroscopic) difference in energy distribution $\triangle E$, a heuristic 
approach to quantum measurement was developed in ref.\cite{Be96}. 
Subsequently, this motivated the present work, which may allow to express
the ideas of~\cite{Be96} within a more appropriate framework. Relevant
questions about the physical nature of detectors are postponed to future
work. This paper is devoted to defining a PSV and the next subsection deals 
with the central issue of constructing a relativistically consistent 
spacetime picture for the reduction process.
\medskip

\subsection{The reduction process in spacetime}

We consider non-overlapping detectors $A, B, C,\, \dots\,,$ each localized 
in a well-defined spacetime region. The basic idea is that these detectors 
perform reductions in a well-defined order, which in the following is 
called {\it reduction order} and indicated by the symbol~$r$. The labelling 
of the detectors may be chosen such that the reduction order 
\begin{equation} \label{rorder}
 r_A < r_B < r_C < \dots 
\end{equation}
holds. If detectors measure at {\it timelike} positions with respect to one 
another, their reduction order is requested to agree with the time order 
of the measurements: $r_A<r_B \Leftrightarrow t_A<t_B$. For detectors at
{\it spacelike} positions the reduction order is still assumed to exist, but 
does no longer correspond to a well-defined time order. We use the reduction 
order to construct a PSV $|\Psi\rangle = |\Psi_W\rangle$ in spacetime. As 
spacelike operators commute, measurable effects are supposed to be independent 
of the chosen reduction order.
\medskip

By assumption, the PSV is initially defined on a LCSH $S_{0+}$:
$|\Psi\rangle = |\Psi(S_{0+})\rangle$ and all reductions happen in the
future (as defined in II.A) of $S_{0+}$. Detector $A$ is first in reduction 
order. Some section of its BLC is in the future of $S_{0+}$ and denoted
$C^{blc}_1$. The next LCSH $S_{1-}$ is obtained by replacing the part of
$S_{0+}$ which is in the past of $C^{blc}_1$ by $C^{blc}_{1-}$. By 
unitary and causal time evolution of law~(a$'$) the PSV becomes defined on
$S_{1-}$: $|\Psi\rangle = |\Psi(S_{1-})\rangle$. This is always possible, 
because the definition of the reduction order makes sure that no reduction 
process can be in the way. For this causal evolution it is sufficient to
know the PSV on the section of $S_{0+}$ which is within the BLC of the
detector's reduction position. 
On $S_1$ detector $A$ performs then the stochastic reduction 
of law~(b$'$) and the PSV becomes defined on $S_{1+}$: $|\Psi\rangle = 
|\Psi(S_{1+})\rangle$. When the initial hypersurface $S_{0+}$ is at 
$t=-\infty$ for all its space points, then $S_1$ is just the BLC of $A$. 
\medskip

Involving detector $B$, the promotion of the PSV from $S_{1+}$ to $S_{2-}$ 
and from there to $S_{2+}$ follows precisely the same scheme. By induction 
we proceed from detector to detector, always promoting the LCSH, with the
PSV defined on it, forward in time: $S_0 < S_1 < S_2 < \dots$ in the 
sense of time-ordering of LCSHs. Should there be a last detector, we can 
promote the PSV by unitary and causal time evolution to $t=+\infty$ for all 
space points. Figure~2 illustrates a situation involving three detectors.
\medskip

Our construction exploits the fact that spacelike operators commute to
attribute each reduction to the action of a local detector. Despite the
fact that the reduction order cannot be verified experimentally, this 
has some physical appeal. In contrast to this the procedure proposed by 
Hellwig and Kraus~\cite{HeKr70} is entirely formal and remains 
inconsistent (see section~III.B) even after our critical assessment
of AA measurements.

\subsection{Aharonov-Albert measurements}

The statement of Aharonov and Albert (AA) is that it is possible to design 
procedures which combine several local interactions to measure some 
non-local property of the quantum system. For instance, they propose two 
arrays of devices, called AA devices henceforth, to measure the non-local 
variable $x_1-x_2$ of two distinguishable particles. These devices interact 
with the quantum system $|\psi_1\rangle\,|\psi_2\rangle$ through the 
Hamiltonian
\begin{equation} \label{AA}
H_{\rm int} = \int di\, (h_1^i+h_2^i) ~~{\rm with}~~
h_j^i = g(t)\,q_j^i\,x_j\,\delta (x_j-x^i_j),\ (j=1,2)\,,
\end{equation}
where $x_j$ is the position of particle $j$, $x_j^i$ is the position of
the $i$th AA device corresponding to particle $j$, and $q_j^i$ is some
internal variable of the device. The coupling $g(t)$ is non-zero only during
a short interval $t_1<t<t_2$, when the devices are switched on. The position
of particle $j$ becomes then coupled to the canonical conjugate momentum
$$ \Pi_j = \int di\ \pi_j^i $$
of array $j$, where $\pi_j^i$ is the canonical conjugate momentum of the 
correspondingly labelled AA device. AA claim that it is possible to
prepare all devices in initial states such that, when separated, they allow
for measurement $x_1-x_2$, but not for measurement of $x_1$, $x_2$ or
$x_1+x_2$, in accordance with
\begin{equation} 
\left[ \Pi_1-\Pi_2 , Q_1+Q_2 \right] = 0 ~~{\rm but}~~ 
\left[ \Pi_j , Q_1+Q_2 \right] \ne 0 ~~{\rm and}~~
\left[ \Pi_1+\Pi_2 , Q_1+Q_2 \right] \ne 0\,.
\end{equation}
See equations (7) to (16) in their paper \cite{AhAl80} for more details.
\medskip

Let us try to combine the AA approach with the concept of measurement developed
in this paper. First, the interactions of nature cannot be turned on and off
at will. They are all supposed to be described by the Lagrangian ${\cal L}$ of 
a fundamental QFT. However, this point is minor. It is perfectly 
reasonable to assume that 
we can prepare AA devices $|\Psi_{AA}^{j,i}\rangle$ such that, through 
preparation of their initial state, their interaction with the quantum
system $|\psi_1\rangle\,|\psi_2\rangle$ is effectively described by
Hamiltonians $H_{\rm int}^{j,i} = (h_1^i+h_2^i)$ with 
$h_j^i$ given by~(\ref{AA}). A second minor point is that in practice 
we can only employ a finite number of such devices, say $i=1,\dots,n$. It 
makes sense to assume that such a finite discretization still gives, 
within certain limitations of accuracy and response efficiency, the 
desired result. Hence, the relevant state vector is
\begin{equation} \label{AAstate}
|\Psi\rangle = |\psi_1\rangle\,|\psi_2\rangle\,\prod_{i=1}^n
|\Psi_{AA}^{1,i}\rangle\,|\Psi_{AA}^{2,i}\rangle
\end{equation}
where the factorization has to be understood in the sense discussed for
equation~(\ref{detector}). In particular, at times when interaction takes
place the expression
\begin{equation} 
|\Psi\rangle = |\psi_1 \prod_{i=1}^n \Psi_{AA}^{1,i}\rangle\
               |\psi_2 \prod_{i=1}^n \Psi_{AA}^{2,i}\rangle
\end{equation}
is appropriate to indicate it, compare equation~(\ref{Psi}).
This product structure of devices limits non-local measurement severely
and cannot be avoided, because every degree of freedom in QFT enters as 
a factor into the state vector. An immediate consequence is that the
non-local operator $\Pi_1 - \Pi_2$ (the terminal points of the two AA
arrays are at spacelike positions) has no representation in terms of
detectors. Devices as imagined in the AA paper cannot be part of the
state vector.
\medskip

One may want to address the question, to where the state vector (\ref{AAstate})
eventually leads us. Neither of the two AA arrays alone can be a detector. 
Namely, by its purpose array $j$ is not allowed to produce a definite result 
for its conjugate momentum $\Pi_j$ and hence a local measurement of the 
position $x_j$ of particle $j$~\cite{classical}.
The only way the AA devices could possibly function is through unitary and
causal interaction with the quantum system $|\psi_1\rangle|\psi_2\rangle$,
then feeding the gained information into a final device $|\Psi_C\rangle$ which 
is a detector. Each of the two AA arrays has to be positioned within the BLC 
of the spacetime point at which detector $C$ performs its measurements in
accordance with the scheme developed in the previous two sections. In the 
rest of this paper we interpret AA devices in this sense (AA-like
devices in the introduction and in the figure captions).
\medskip

Not only our notion of measurement is distinct from AA, but also our notion
of a PSV. As element in Fock space our PSV exist only on certain spacelike
hyperplanes (never cross reduction lines!). However, where it does not exist
it cannot be monitored. An example is given in section~III.A.  
\medskip

\section{Examples}

In the next subsection we deal with detection of a single, charged 
particle by detectors $A$ and $B$ at spacelike positions. In addition 
it is shown why non-local monitoring, involving two AA devices and
a third detector $C$ fails.
\medskip

In the other subsections we work out various examples of spacelike 
measurements
on states made from distinguishable spin~${1\over 2}$ particles. 
Detectors are denoted by $A$, $B$ and $C$ and assumed to act at spacelike
positions in reduction order $r_A<r_B<r_C$ (or in a different reduction
order when explicitly stated). Each detector can measure the spin of an 
incoming particle with respect to one of $n$ different axis $\hat{r}_i$, 
$i=1,\dots,n$. 
As the particles are distinguishable, we can (and do) label them by the 
detectors which eventually measure them. The following equation defines 
then eigenstates with respect to the axis $\hat{r}_i$ as measured by 
detector $A$
\begin{equation}
 L^A_i\, |a^{i+}\rangle = {1\over 2}\,|a^{i+}\rangle ~~{\rm and}~~ 
 L^A_i\, |a^{i-}\rangle = - {1\over 2}\, |a^{i-}\rangle
\end{equation}
where the superscripts $+$ and $-$ label the $+{1\over 2}$ and 
$-{1\over 2}$ eigenstates, respectively. The same notation is used for 
the other detectors, replacing $A$ by $B$ or $C$ (and $a$ by $b$ or $c$).

\subsection{Split charged particle}

We consider a single charged particle which has been split
by a magnetic field into a superposition of two localized waves
$$ |\psi\rangle = |\psi_a\rangle + |\psi_b\rangle $$
where $|\psi_a\rangle$ heads towards detector $A$ and $|\psi_b\rangle$
towards detector $B$ along the lines indicated in figure~3~\cite{f5}.
Each detector will measure simply whether a charged particle comes 
in or not. The initial LCSH $S_0$ is chosen to be
at $t=-\infty$, out of the picture. AA devices are positioned on 
spacetime worldlines $AA_1$ and $AA_2$ of this figure (as they are not 
measurement devices the same kind of lines as for QOs are used). 
In the present example we assume that the charged particle is scalar
or an electron with given spin. The no-cloning theorem~\cite{nct} does
then allow to assume that an AA device 
is able to make a copy of the state it encounters 
and to send this copy to the detector~$C$. Without bothering about 
the experimental feasibility, this seems to be the
most favorable assumption about the eventual performance of such
devices. The copy will be denoted by $|\psi^i_c\rangle\ (i=1,2)$
when originating from AA device $AA_i$. Detector $C$ performs the
same kind of measurement as detectors $A$ and $B$ do and allows to
trace the direction from where the charged particle comes.
\medskip

Detector $A$ reduces the state vector on the LCSH $S_1$. On the $S_{1-}$ 
side of this surface the PSV, obtained from $S_{0+}$ by unitary and causal 
time evolution, reads
\begin{equation} \label{S1-a}
|\Psi (S_{1-}) \rangle =
 (|\psi_a\rangle\,|\psi_c^1\rangle + |\psi_b\rangle)\,
 |\Psi^1_{AA}\rangle\,|\Psi^2_{AA}\rangle\,
 |\Psi_A\rangle\,|\Psi_B\rangle\,|\Psi_C\rangle\,,
\end{equation}
where the state vector includes detectors and AA devices. $AA_1$ has
created the copy $|\psi^1_c\rangle$ of $|\psi_a\rangle$. The reduction
(100\% efficiency assumed) by detector $A$ transforms $|\Psi\rangle$ 
either into
\begin{equation} \label{S1a}
|\Psi (S_{1+}) \rangle = |\psi_c^1\rangle\,|\Psi^1_{AA}\rangle\,
|\Psi^2_{AA}\rangle\,|\Psi'_A\rangle\,|\Psi_B\rangle\,|\Psi_C\rangle\,,
\end{equation}
where the prime on $|\Psi'_A\rangle$ indicates that a charged particle
has been detected, or into
\begin{equation} \label{S1b}
|\Psi (S_{1+}) \rangle = |\psi_b\rangle\,|\Psi^1_{AA}\rangle\,
|\Psi^2_{AA}\rangle\,|\Psi_A\rangle\,|\Psi_B\rangle\,|\Psi_C\rangle\,.
\end{equation}
when no particle is detected by $A$. In either case $|\Psi\rangle$ is 
then defined on the LCSH $S_{1+}$. Continuing with~(\ref{S1a}), the
final state state, defined on the spacelike surface $S_4$ of 
figure~3, will be
\begin{equation} \label{fina}
|\Psi (S_4)\rangle = |\Psi^1_{AA}\rangle\,|\Psi^2_{AA}
\rangle\,|\Psi'_A\rangle\,|\Psi_B\rangle\,|\Psi'_C\rangle\
\end{equation} 
{\it i.e.} a charged particle has now also been detected by detector
$C$ in a measurement which is not a reduction. As the picture for this
case is rather obvious, it is not drawn in figure~3, instead the next 
one is depicted.  Continuing 
with~(\ref{S1b}) the second AA device has an opportunity to act 
and on the spacelike surface $S_{2-}$ of figure~3 the state becomes
$$ |\Psi (S_{2-})\rangle = |\psi_b\rangle\,|\psi^2_c\rangle\,
|\Psi^1_{AA}\rangle\,|\Psi^2_{AA}\rangle\,
|\Psi_A\rangle\,|\Psi_B\rangle\,|\Psi_C\rangle\,. $$
After measurements by $B$ and $C$ which are both no reductions the 
final state becomes
\begin{equation} \label{finb}
 |\Psi (S_4)\rangle = 
 |\Psi^1_{AA}\rangle\,|\Psi^2_{AA}
 \rangle\,|\Psi_A\rangle\,|\Psi'_B\rangle\,|\Psi'_C\rangle\,. 
\end{equation} 
In (\ref{fina}) as well as in (\ref{finb}) the detector $C$ has measured a 
charge, but it {\it cannot} be interpreted as measuring a non-local charge.
\medskip

Let us push this example a bit further and assume now the reduction 
order $r_C<r_B<r_A$. The BLC of detector $C$ defines now the reduction 
surface $S_1$ as indicated in figure~4. On the $S_{1-}$ side the PSV is
\begin{equation} \label{S1-c}
|\Psi\rangle = (|\psi_a\rangle\,|\psi_c^1\rangle + |\psi_b\rangle\,
|\psi_c^2\rangle)\,|\Psi^1_{AA}\rangle\,|\Psi^2_{AA}\rangle\,
|\Psi_A\rangle\,|\Psi_B\rangle\,|\Psi_C\rangle
\end{equation}
where both $AA$ devices have now done their work. Detector $C$
reduces the state to either
\begin{equation} \label{S1ca}
|\Psi(S_{1+})\rangle = |\psi_a\rangle\,|\Psi^1_{AA}\rangle\,|\Psi^2_{AA}
\rangle\,|\Psi_A\rangle\,|\Psi_B\rangle\,|\Psi'_C\rangle\,,
\end{equation}
or
\begin{equation} \label{S1cb}
|\Psi(S_{1+})\rangle = |\psi_b\rangle\,|\Psi^1_{AA}\rangle\,
|\Psi^2_{AA}\rangle\,|\Psi_A\rangle\,|\Psi_B\rangle\,|\Psi'_C\rangle\,.
\end{equation}
The final PSVs are identical to those obtained before: Equation (\ref{S1ca})
leads to (\ref{fina}) and equation~(\ref{S1cb}) to (\ref{finb}). The latter 
one is drawn in figure~4. As observed
in~\cite{AhAl80} the charge is not conserved on all spacelike hyperplanes.
As charge is a superselection rule, this means on such hyperplanes a PSV as
vector in Fock space does not exist. Our example enlightens that these 
non-existing states cannot be monitored and our PSV evolution remains 
empirically satisfactory.
\medskip

\subsection{Singlet decay}

As in the previous subsection the $S_0$ LCSH is moved to $t=-\infty$.
Including the detectors $A$ and $B$, the singlet state reads
\begin{equation} \label{singlet}
|\Psi (S_{1-})\rangle = 
2^{-{1\over 2}}\,\{ |a^{k-}\rangle\,|b^{k+}\rangle
- |a^{k+}\rangle\, |b^{k-}\rangle \}\,|\Psi_A\rangle\,|\Psi_B\rangle .
\end{equation}
Assume that $A$ measures with respect to axis $\hat{r}_i$ and that
the result is $i+$. The corresponding reduction reads
$$ 2^{-{1\over 2}}\,\{ |a^{k-}\,\Psi_A\rangle\,|b^{k+}\rangle 
 - |a^{k+}\,\Psi_A\rangle\, |b^{k-}\rangle \}\,|\Psi_B\rangle \to
   \{ \langle a^{i+}|a^{k-} \rangle\,|b^{k+}\rangle
 - \langle a^{i+} |a^{k+} \rangle\,|b^{k-}\rangle \}\,
   |\Psi_A^{'i+}\,|\Psi_B\rangle $$
It is easy to work out the matrix elements 
$\langle a^{i+} | a^{k-} \rangle = -\sin (\theta/2)$ and
$\langle a^{i+} | a^{k+} \rangle = \cos (\theta/2)$, where $\theta$ is
the angle between axis $i$ and axis $k$. Further, $|b^{i-}\rangle =
- \sin (\theta/2) |b^{k+} \rangle + \cos (\theta/2) |b^{k-} \rangle$.
Hence, detector $A$ reduces the state to
\begin{equation} \label{sredA}
|\Psi(S_{1+})\rangle =
-|b^{i-}\rangle\,|\Psi_A^{'i+}\rangle\,|\Psi_B\rangle\,,  
\end{equation}
see figure~5 for the location of $S_1$, which cuts through the 
worldline of particle $|b\rangle$ before it reaches detector $B$.
Equation~(\ref{sredA}) still holds on $S_{2-}$. Assume that $B$ 
measures with respect to axis $\hat{r}_j$ and that the result
is $j+$. The final reduction simply reads
\begin{equation} \label{sredB}
- |\Psi_A^{'i+}\rangle\, |b^{-i}\Psi_B\rangle \to
 |\Psi(S_{2+})\rangle = 
- |\Psi_A^{'i+}\rangle\, |\Psi_B^{'j+}\rangle\,.
\end{equation}
Note that the reduction performed by $B$ is entirely local, 
affecting the PSV only in the immediate neighbourhood of $B$.
Given axis $\hat{r}_i$ for $A$ and $\hat{r}_j$ for $B$, let us 
find the QT probability $P(a^{i+},b^{j+})$ that $i+$ is 
measured by $A$ and $j+$ by $B$. For the reduction step from 
(\ref{singlet}) to (\ref{sredA}) the probability is 
\begin{equation} \label{prob1}
\left| \langle\Psi(S_{1+})|\Psi(S_{1-})\rangle \right|^2 = 
\left| 2^{-{1\over 2}}\,\left[ (\langle a^{i+}|a^{k-} \rangle)^2 +
(\langle a^{i+} |a^{k+} \rangle)^2 \right] \right|^2 = {1\over 2}
\end{equation}
and for the step (\ref{sredB}) it is
$|\langle b^{j+} | b^{i-}\rangle|^2$. Together this implies 
\begin{equation} 
P(a^{i+},b^{j+})={1\over 2}\,|\langle b^{j+}|b^{i-}\rangle|^2\,.
\end{equation}
Note that for $j+=i-$ the matrix element becomes 
$|\langle b^{i-} | b^{i-}\rangle|^2 = 1$: Detector $B$ still
measures, but no reduction takes place. Finally, the reader can 
easily verify that the same probability $P(a^{i+},b^{j+})$, namely
${1\over 2} |\langle a^{j+} | a^{i-}\rangle|^2$, is obtained when
$B$ measures $j+$ first in reduction order, and $A$ measures 
then~$i+$.
\medskip

We compare now with the construction by Hellwig and Kraus~\cite{HeKr70}.
As in figure~3 of their paper, regions~1 to~4 are defined in our figure~6.
Positions of the detectors $A$ and $B$ are identical with those in
figure~5, but now both BLCs are drawn. We ignore detector $C$ at the
moment and construct a state on a spacelike 
hypersurface within one region of figure~6 by crossing all worldline 
available there, but never leaving the region. For the measurement process 
described above, we obtain the following results:

\begin{itemize}
\item Equation (\ref{singlet}) in region 1.
\item $+|a^{j-}\rangle\,|\Psi_A\rangle$ in region 2 due to measurement at $B$.
\item $-|b^{i-}\rangle\,|\Psi_B\rangle$ in region 3 due to measurement at $A$, 
      compare equation~(\ref{sredA}).
\item $-|\Psi_A^{'i+}\rangle\,|\Psi_B^{'j+}\rangle$ in region 4, compare
      equation~(\ref{sredB}).
\end{itemize}

The major difference to the construction of this paper is that the
assignment in the regions~2 and~3 is a formal one, defining the results
as if both detectors $A$ and $B$ would have carried out their reductions.
Serious consequences are implied. 
Let us introduce AA devices $AA_1$ and $AA_2$ as in the 
previous subsection, see figure~3. They create the state 
$+|a^{j-}\rangle\,|c_1^{j-}\rangle\,|\Psi_A\rangle$ in region~2 and
$-|b^{i-}\rangle\,|c_2^{i-}\rangle\,|\Psi_B\rangle$ in region~3.
This amounts to sending the copy state 
$|c_1^{j-}\rangle\,|c_2^{i-}\rangle$ to detector $C$, resulting in the
conditional probability $P(c_1^{j-},c_2^{i-})=1$ in case that detector $C$
measures the spin of particle $|c_1\rangle$ with respect to the 
$\hat{r}_j$ axis and the spin of particle $|c_2\rangle$ with respect to
the $\hat{r}_i$ axis. This probability is in disagreement with QT 
({\it i.e.} false).
\medskip

In contrast, the reduction process of this paper stays correct by the 
simple fact that $C$ performs just another spacelike, local measurement 
({\it i.e} commutes with $A$ and $B$).
Straightforward but lengthy algebra allows to verify explicitly that 
all reduction orders lead to the same result. For the remainder of this 
subsection, we follow our standard order $r_A<r_B<r_C$ and include 
$AA$~devices.
\medskip

At the positions indicated in figure~3 we imagine the AA devices $AA_1$ 
and $AA_2$ in figure~5. For notational convenience, we do no longer 
include detectors and AA devices explicitly. After passing $AA_1$ the 
state vector becomes then  
\begin{equation}
 |\psi(S_{1-})\rangle = 2^{-{1\over 2}}\,\{
   |a^{k-}\rangle |b^{k+}\rangle |c_1^{k-}\rangle
 - |a^{k+}\rangle |b^{k-}\rangle |c_1^{k+}\rangle \}\,,
\end{equation}
When reduction by detector $A$ has produced the measurement result 
$i+$ we have
\begin{equation}
 |\psi(S_{1+})\rangle = 
   \langle a^{i+}|a^{k-}\rangle |b^{k+}\rangle |c_1^{k-}\rangle
 - \langle a^{i+}|a^{k+}\rangle |b^{k-}\rangle |c_1^{k+}\rangle
\end{equation}
and this happens with with probability (\ref{prob1}). After passing 
$AA_2$ the state vector becomes
\begin{equation} \label{newent}
 |\psi (S_{2-}) \rangle = 
   \langle a^{i+}|a^{k-}\rangle   |b^{k+}\rangle 
                 |c_1^{k-}\rangle |c_2^{k+}\rangle
 - \langle a^{i+}|a^{k+}\rangle |b^{k-}\rangle 
                 |c_1^{k+}\rangle |c_2^{k-}\rangle\,.
\end{equation}
Remarkable with this equation are two points: (1) The same result 
is obtained when passing $AA_2$ proceeds the reduction by detector $A$. 
(2)~Due to the locality  of their interaction, the AA devices are unable 
to copy ({\it i.e.} measure non-locally) the fully entangled 
state~(\ref{singlet}), which would be 
$$\sim\ \{ |c_1^{k-}\rangle\,
|c_2^{k+}\rangle-|c_1^{k+}\rangle\,|c_2^{k-}\rangle \}\,.$$
{\it I.e.} this expression would have to show up as a simple overall 
factor. Instead, the copying process has 
produced a new entanglement. The device has no ability to monitor non-locally
the spin zero content of the initial state, as envisioned in~\cite{AhAl80}.
Still, this new entanglement is non-local enough to produce an inconsistency 
for the prescription proposed by Hellwig and Kraus. Namely, the reduction
by $B$ for the measurement $j+$ yields
\begin{equation}
 |\psi (S_{2+}) \rangle = const\, \{ 
   \langle a^{i+}|a^{k-}\rangle   
   \langle b^{j+}|b^{k+}\rangle 
                 |c_1^{k-}\rangle |c_2^{k+}\rangle
 - \langle a^{i+}|a^{k+}\rangle 
   \langle b^{j+}|b^{k-}\rangle 
                 |c_1^{k+}\rangle |c_2^{k-}\rangle \}
\end{equation}
where the constant is determined by the normalization condition 
$\langle \psi | \psi \rangle =1$. The probability for this to happen is
a bit lengthy and spared the reader, because it is not really of 
importance. Important is, once it happened the conditional probability
that detector $C$ finds $j-$ for particle $|c_1\rangle$ and $i-$ for
particle $|c_2\rangle$ is in general less than one, as follows from
calculating $P(c_1^{j-},c_2^{i-}) = |\langle c_1^{j-}|\, 
\langle c_2^{i-}\,|\,\psi (S_{2+}) \rangle |^2$.

\subsection{The Greenberger--Horne--Zeilinger state}

A GHZ state~\cite{GHZ89} is a state of three distinguishable 
spin~${1\over 2}$ particles which gained some popularity~\cite{Me91} as 
a theoretical example for which local realistic theories and QT differ 
in 100\% of the results for some measurements. Here our interest in it 
is limited to 
illustrating reduction for a case involving three detectors. For the
following discussion we define $\hat{r}_1=\hat{x}$, $\hat{r}_2=\hat{y}$
and $\hat{r}_3=\hat{z}$ and do not consider any other axis. Reduction
for our standard order is depicted in figure~7. Including the detectors, 
one of the eight GHZ states reads~\cite{Me91}
\begin{equation} \label{GHZ}
|\Psi (S_{1-})\rangle = 2^{-{1\over 2}}\, \{ 
   |a^{3+}\rangle\,|b^{3+}\rangle\,|c^{3+}\rangle
 - |a^{3-}\rangle\,|b^{3-}\rangle\,|c^{3-}\rangle \}\,
|\Psi_A\rangle\,|\Psi_B\rangle\,|\Psi_C\rangle\, .
\end{equation}
Assume that $A$ measures the $L_x$ spin component and that the result 
is $+{1\over 2}$. The corresponding reduction is
\begin{equation} \label{GredA}
 |\Psi(S_{1-})\rangle \to |\Psi(S_{1+})\rangle = 2^{-{1\over 2}}\,\{
 |b^{3+}\rangle\,|c^{3+}\rangle\ - |b^{3-}\rangle\,|c^{3-}\rangle \}\,
 |\Psi_A^{'1+}\rangle\,|\Psi_B\rangle\,|\Psi_C\rangle \}\, . 
\end{equation}
Next, let $B$ measures $L_y$ and the result is assumed to be 
$+{1\over 2}$ again. Then the reduction on $S_2$ of figure~7 is
\begin{equation} \label{GredB}
 |\Psi(S_{2-})\rangle \to |\Psi(S_{2+})\rangle = \{
 |c^{3+}\rangle\ + i\,|c^{3-}\rangle \}\,
 |\Psi_A^{'1+}\rangle\,|\Psi_B^{'2+}\rangle\,|\Psi_C\rangle \} =
 2^{-{1\over 2}}\,|c^{2+}\rangle\,
 |\Psi_A^{'1+}\rangle\,|\Psi_B^{'2+}\rangle\,|\Psi_C\rangle \}\, .
\end{equation}
Reductions (\ref{GredA}) and (\ref{GredB}) are both non-local. Note that
the geometry of figure~7 is chosen such that the change affecting particle
$|c\rangle$ happens in a region where $S_2$ falls, up to the proper definition
of limits, on $S_1$. This is the reason why different line symbols are used
to indicate the continuation of the $|b\rangle$ and $|c\rangle$ 
worldlines after reduction. The reduction for $|b\rangle$ is on $S_1$, whereas
$|c\rangle$ undergoes two reductions: on $S_1$ and $S_2$ in a region where
these surfaces fall on top of one another. The final reduction by detector 
$C$ is local and when detector $C$ measures $L_y$ the result is with
certainty $+{1\over 2}$, {\it i.e.} a measurement without reduction.

\section{Conclusions}

The central point of this paper is that it is possible to understand 
reduction in QFT as a {\it fundamental physical process}. Each reduction
is performed {\it locally} by a detector, whereas the consequences can
be {\it global}. Essential for such an interpretation of the reduction 
process is the existence of a consistent spacetime picture. This was 
shown by explicit construction of a PSV. 
In contrast to the non-relativistic limit the PSV can no longer be 
monitored, because the order in which detectors at spacelike separations 
perform their reductions cannot be verified experimentally. Therefore,
the developed spacetime picture has not to be taken literally. 

Potentially, our framework is a building block of a theory of reduction.
It opens the door towards investigating questions like: Which 
conglomerates of matter have under which circumstances the ability 
to perform reductions ({\it i.e.} can act as detectors)?  Are there
frequency laws of reduction? These issues will be addressed in future 
work. Preliminary thoughts exist~\cite{Be96} and, actually, motivated 
the present investigation.
The goal is to formulate reduction as a stochastic process, which goes 
on independently of whether measurements are performed or not, such that
the measurement process becomes explained. Interestingly, there can then
be experimental implications beyond standard QT which allow to verify
such a stochastic process.

\vskip.1in

\noindent {\bf Acknowledgement:} This work has been partially supported 
by the Department of Energy under contract number DE-FG02-97ER41022.


\bigskip


\begin{thebibliography}{99}

\bibitem{Bl67} I. Bloch, Phys. Rev. {\bf 156} (1967) 1377.

\bibitem{exp} A. Aspect, J. Dalibert, and G. Roger, Phys. Rev. Lett. {\bf 49}
(1982) 1804; P.R. Tapster, J.G. Rarity, and P.C.M. Owens, Phys. Rev. Lett.
{\bf 73} (1994) 1923; W. Tittel, J. Brendel, B. Gisin, T. Herzog, H.Zbinden
and N. Gisin, quant-ph/9707042.

\bibitem{Be64} J.S. Bell, Physica {\bf 1} (1964) 195.

\bibitem{HeKr70} K.-E. Hellwig and K. Kraus, Phys. Rev. {\bf D1} (1970) 566.

\bibitem{AhAl80} Y. Aharonov and D. Z. Albert, Phys. Rev. {\bf D21} (1980) 3316.

\bibitem{GHZ89} D.M. Greenberger, M. Horne and A. Zeilinger, in {\it Bell's
Theorem, Quantum Theory and Conceptions of the Universe}, edited by
M.~Kafatos, Kluwer Academic, Dordrecht, 1989, pp.69-72.

\bibitem{Me91} N.D. Mermin, Am. J. Phys. {\bf 58} (1991) 731.

\bibitem{time} To be definite the proper time of the detector may be chosen.
By construction a detector is massive and has a well-defined rest frame.

\bibitem{Be96} B.A. Berg, hep-ph/9609232 (revised July 1998).

\bibitem{classical} Notably, a classical interpretation of the AA devices is 
also ruled out by this requirement, because an instrument~$j$ of classical
physics would have a well defined conjugate momentum $\Pi_j$ at all times.

\bibitem{f5} In case of difficulties consider first the simpler situation 
of figure~5 in the next subsection. 

\bibitem{nct} D. Diek, Phys. Lett. {\bf 92A} (1982) 271;
W.K. Wootters and W.H. Zurek, Nature {\bf 299} (1982) 802.

\end{thebibliography}
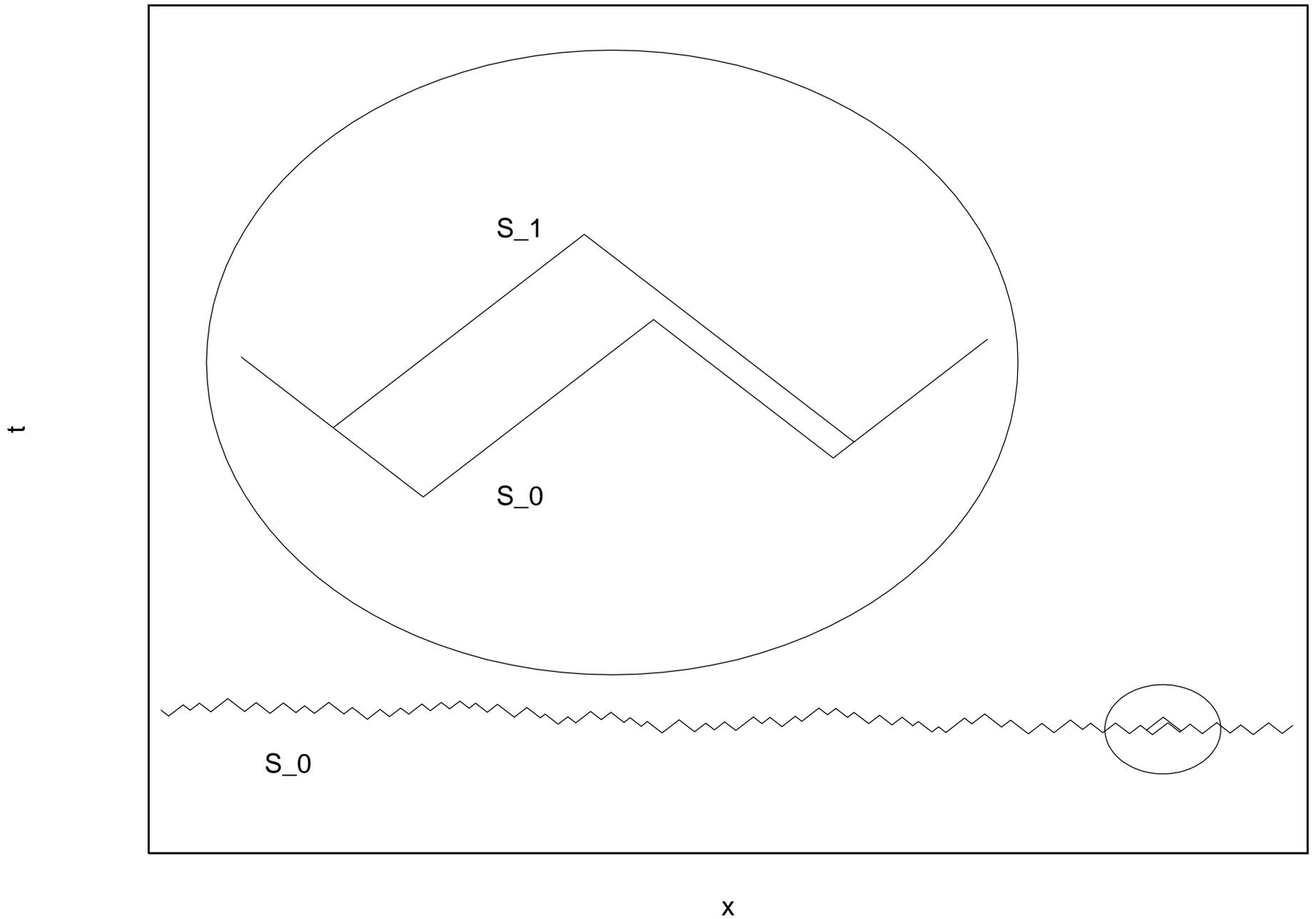
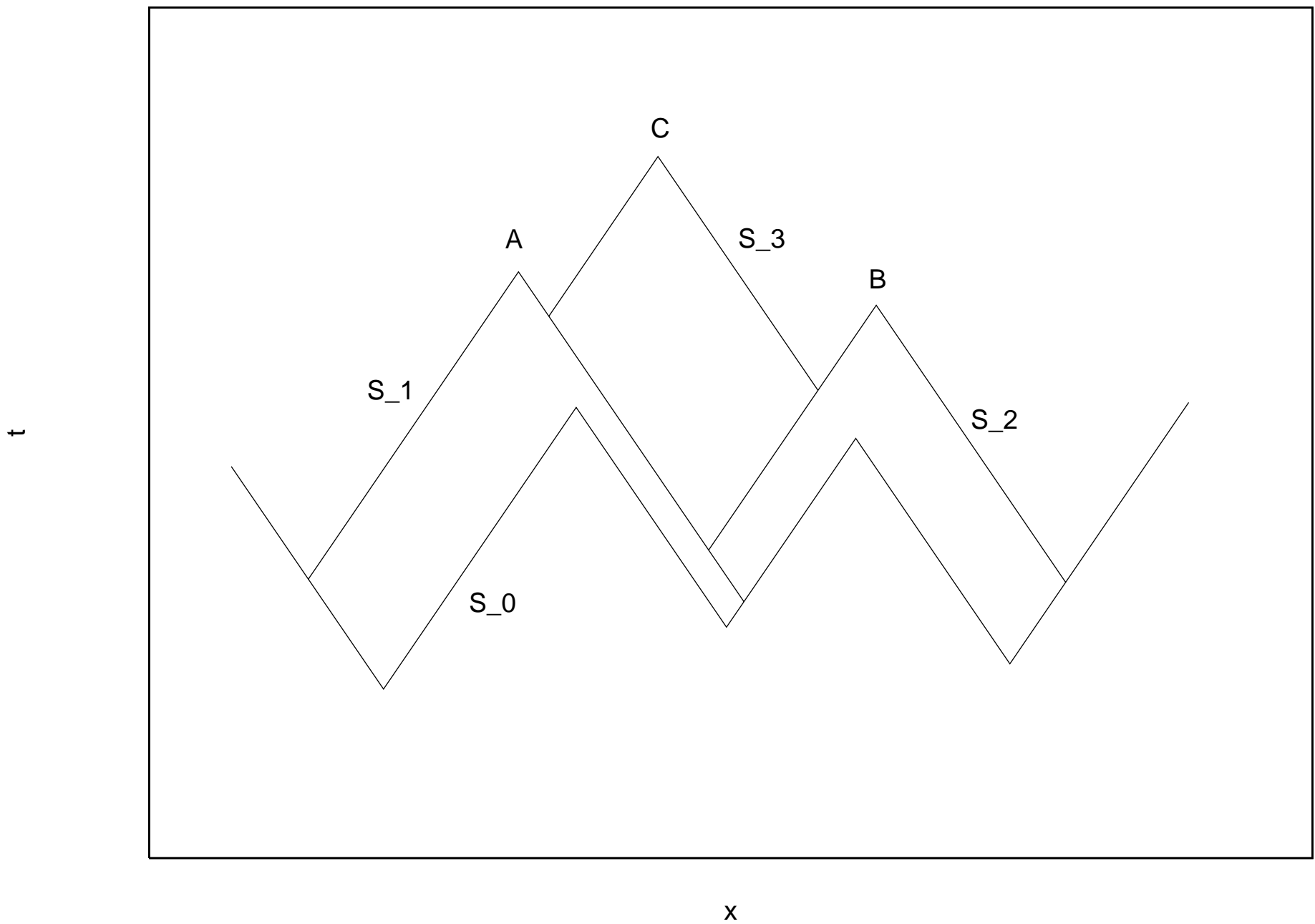
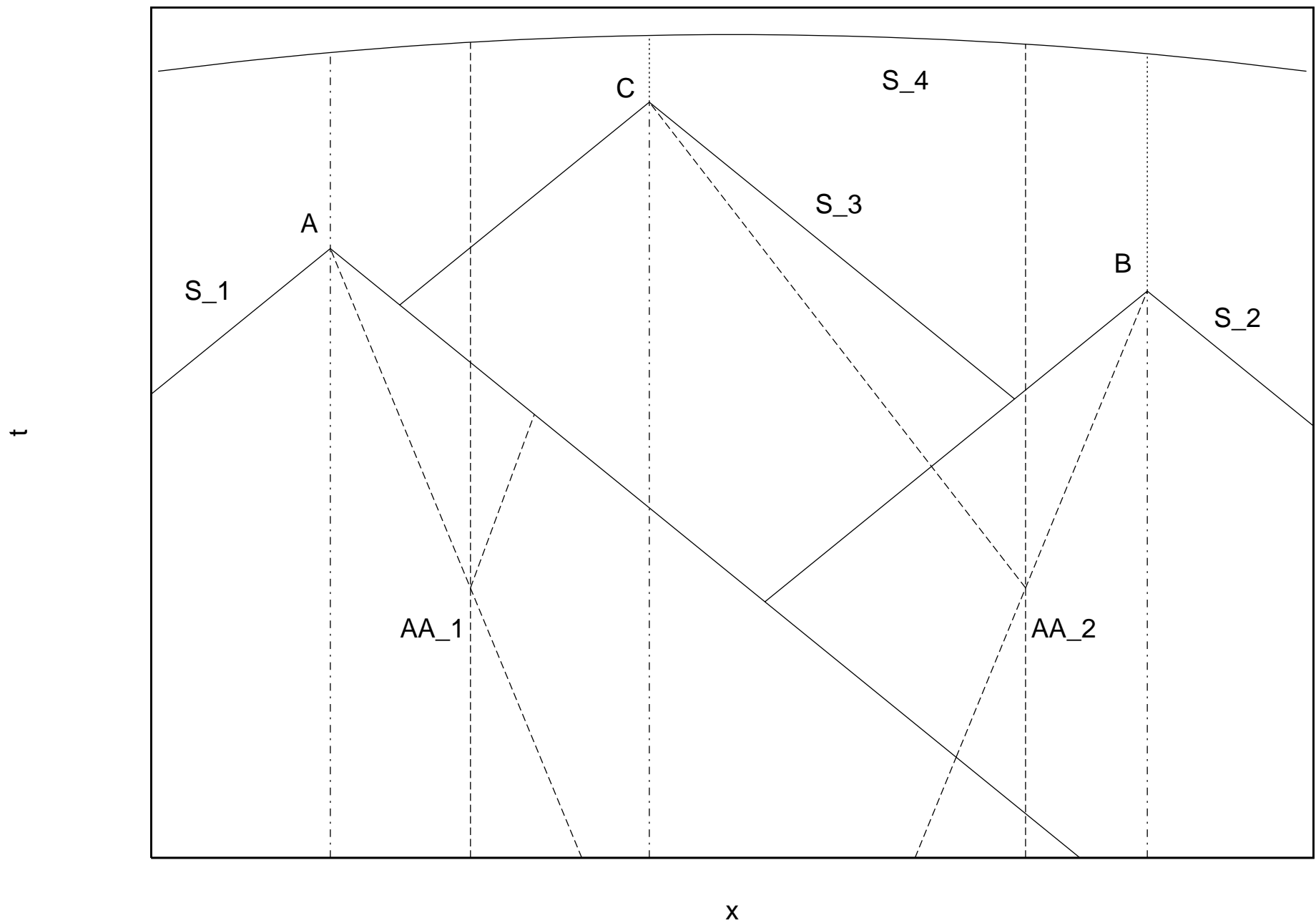
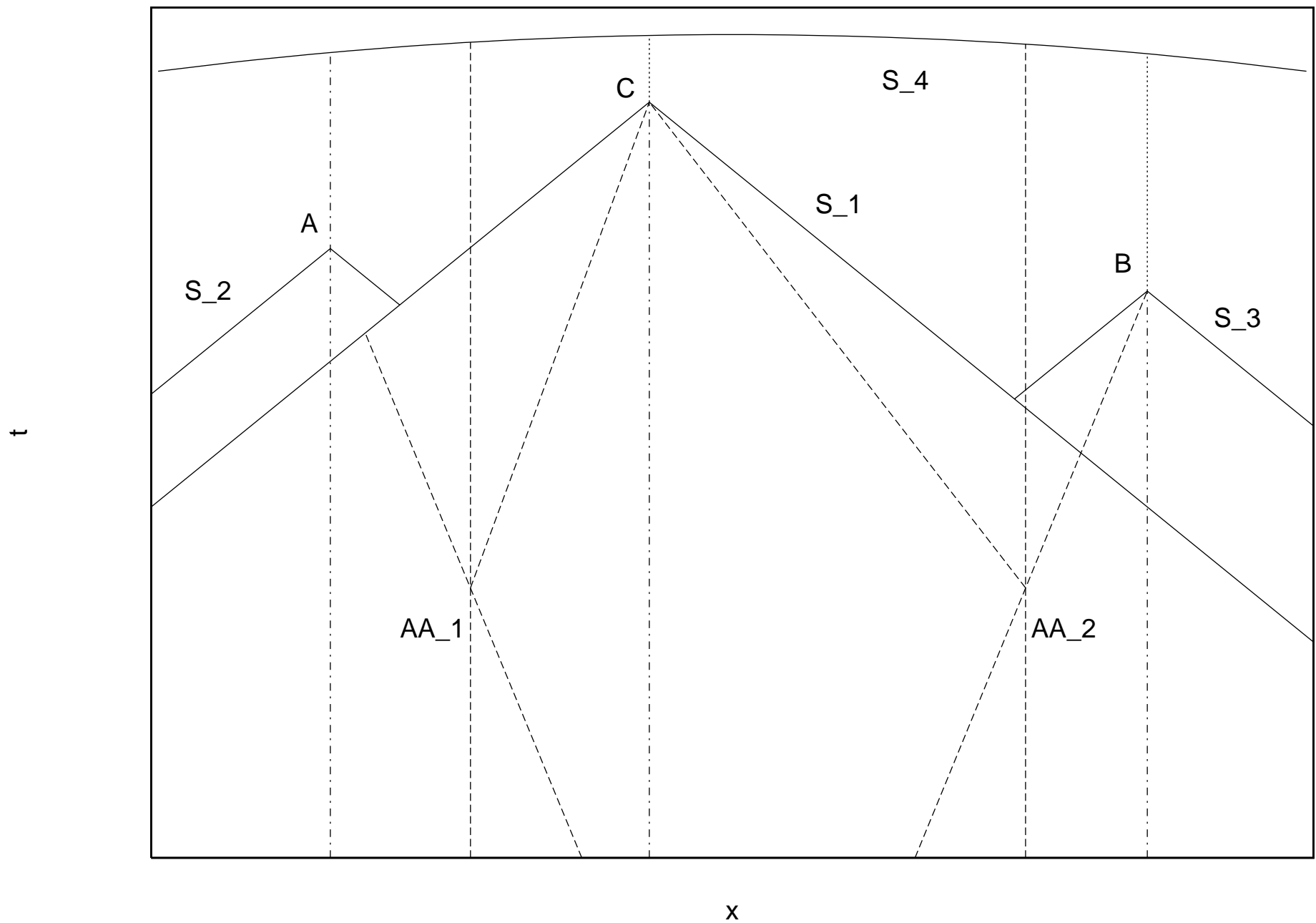
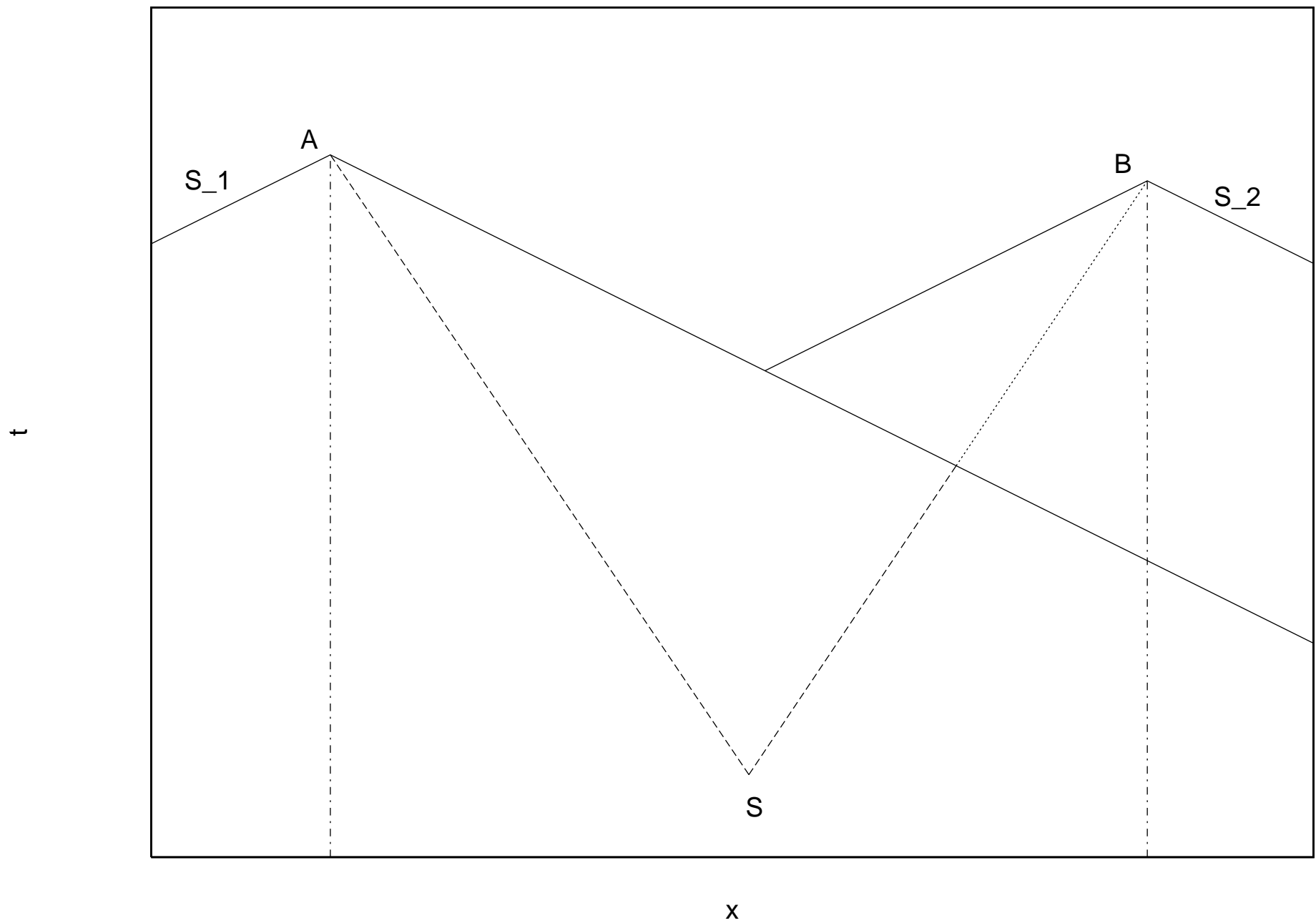
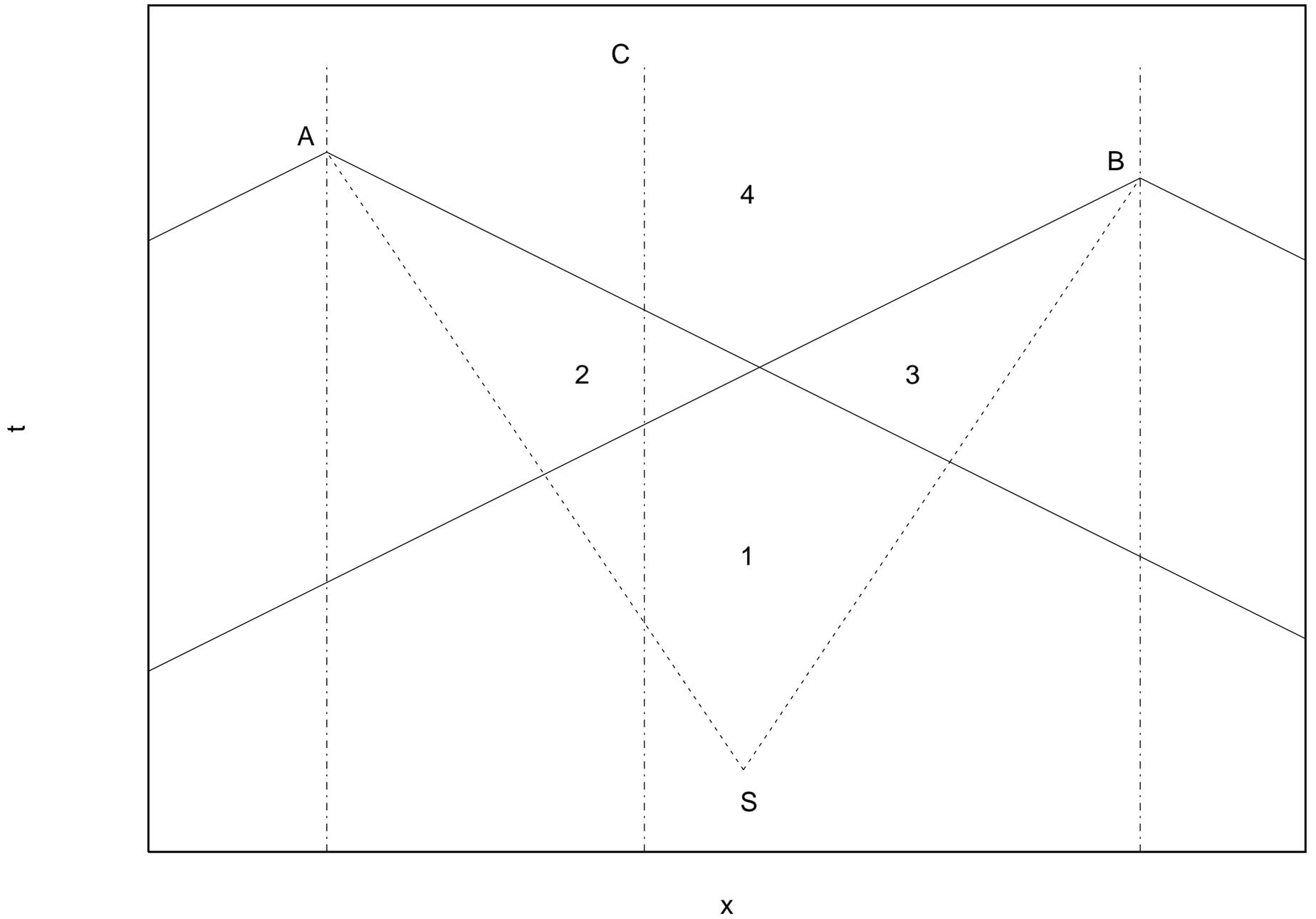
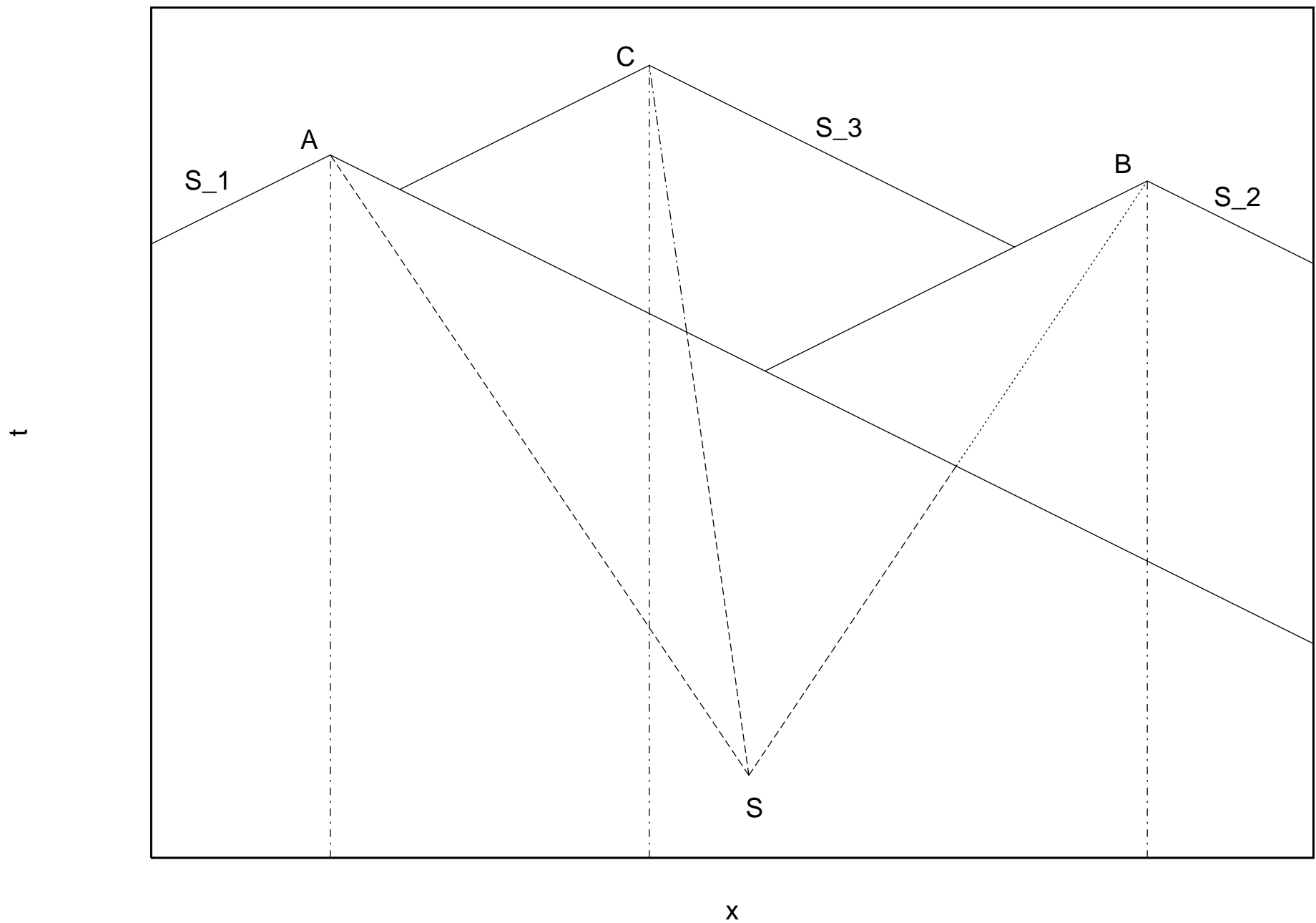
\end{document}